# 基于数据分析和非线性规划的蔬菜类商品自动定价与补货策略

马铭璞

（天津大学管理与经济学部，天津 300072）

## 摘　要

　　在生鲜商超领域，一般而言，蔬菜类商品具有相对有限的保鲜期，其品质随销售时间的推移而逐渐下降。大多数蔬菜品种如果未能在当日售出，次日将难以再次销售。因此，商超通常根据商品的历史销售数据和需求情况，每日进行定量补货操作。蔬菜的定价一般采用"成本加成定价"方法，商超通常会对因运输损耗和品质下降而影响的商品进行打折销售。在这一背景下，可靠的市场需求分析显得尤为重要，因为它直接关系到补货和定价决策的制定。由于商超的销售空间有限，因此合理的销售组合变得至关重要。本文基于所提供的蔬菜品类的商品信息、蔬菜的销售流水明细数据、蔬菜类商品的批发价格和蔬菜类商品的近期损耗率等数据，首先利用数据分析和可视化技术分析了蔬菜各品类及单品销售量的分布规律及相互关系，其次，构建蔬菜品类的销售总量与成本加成定价的函数关系，利用 ARIMA 模型对蔬菜品类的未来批发价格进行了预测，基于此建立销售利润函数与约束条件，进一步构建非线性规划模型并进行求解，给出了商超各蔬菜品类未来一周的日补货总量和定价策略。进一步地，我们结合商超实际销售的情况与要求，对利润函数和约束条件进行了优化，在尽量满足市场对各品类蔬菜商品的需求的前提下，给出了 7 月 1 日的单品补货量和定价策略，以使得商超收益最大化。最后，为了更好地制定蔬菜商品的补货和定价决策，我们对商超需要采集的数据进行了讨论与展望，并具体分析了所收集的数据如何应用于上述问题。

**关键词**：时序预测、非线性规划、生鲜产品管理、库存优化、ARIMA 预测、数据驱动决策、零售利润最大化、蔬菜保鲜期管理

# Automatic Pricing and Replenishment Strategies for Vegetable Products Based on Data Analysis and Nonlinear Programming

Mingpu Ma[*]

[*] College of Management and Economics, Tianjin University, Tianjin 300072, China

E-mail: m125_0409@tju.edu.cn

## ABSTRACT

In the field of fresh produce retail, vegetables generally have a relatively limited shelf life, and their quality deteriorates with time. Most vegetable varieties, if not sold on the day of delivery, become difficult to sell the following day. Therefore, retailers usually perform daily quantitative replenishment based on historical sales data and demand conditions. Vegetable pricing typically uses a "cost-plus pricing" method, with retailers often discounting products affected by transportation loss and quality decline. In this context, reliable market demand analysis is crucial as it directly impacts replenishment and pricing decisions. Given the limited retail space, a rational sales mix becomes essential. This paper first uses data analysis and visualization techniques to examine the distribution patterns and interrelationships of vegetable sales quantities by category and individual item, based on provided data on vegetable types, sales records, wholesale prices, and recent loss rates. Next, it constructs a functional relationship between total sales volume and cost-plus pricing for vegetable categories, forecasts future wholesale prices using the ARIMA model, and establishes a sales profit function and constraints. A nonlinear programming model is then developed and solved to provide daily replenishment quantities and pricing strategies for each vegetable category for the upcoming week. Further, we optimize the profit function and constraints based on the actual sales conditions and requirements, providing replenishment quantities and pricing strategies for individual items on July 1 to maximize retail profit. Finally, to better formulate replenishment and pricing decisions for vegetable products, we discuss and forecast the data that retailers need to collect and analyses how the collected data can be applied to the above issues.

# 目 录

# 一、 问题分析

在生鲜超市管理领域，涉及一系列复杂问题，包括供应链管理、定价策略以及市场需求分析等方面。以蔬菜类商品为案例，这些商品在生鲜商超中具有较短的保鲜期，其品质随销售时间的推移而逐渐降低。多数蔬菜品种如当日未售出，次日将难以再次销售。因此，商超通常会根据各商品的历史销售数据和需求情况，每日进行补货。由于商超销售的蔬菜品种繁多且产地多样化，而且蔬菜的采购通常在凌晨 3:00 至 4:00 进行，这使得商家必须在不完全了解特定单品和采购价格的情况下，制定当日各蔬菜品类的补货决策。一般而言，蔬菜的定价采用"成本加成定价"方法，商超通常会对由于运输损耗和品质下降而受影响的商品进行打折销售。可靠的市场需求分析在补货和定价决策方面具有特殊重要性。从需求方面来看，蔬菜类商品的销售量通常与时间存在一定的相关性；从供给方面来看，4 月至 10 月期间蔬菜供应的品种相对较多，但由于商超的销售空间受限，因此制定合理的销售组合变得至关重要。因此，本文需解决如下问题：

（1）蔬菜类商品不同品类或不同单品之间可能存在一定的关联关系，基于所提供的数据，分析蔬菜各品类及单品销售量的分布规律与相互关系。

（2）考虑商超以品类为单位做补货计划，分析各蔬菜品类的销售总量与成本加成定价的关系，并给出各蔬菜品类未来一周（2023 年 7 月 1 到 7 日）的日补货总量和定价策略，使得商超收益最大。

（3）因蔬菜类商品的销售空间有限，商超希望进一步制定单品补货计划，要求可售单品总数控制在 27 到 33 个，且各单品订购量满足最小陈列量 2.5 千克的要求。根据 2023 年 6 月 24 到 30 日的可售品种，给出 7 月 1 日单品补货量和定价策略，在尽量满足市场对各品类蔬菜商品需求的前提下，使得商超收益最大。

（4）为了更好地制定蔬菜商品的补货和定价决策，分析商超还需要采集哪些相关数据并如何运用。

## 二、 模型假设

为了构建相关数学模型，我们给出一些模型假设：

1. 数据真实性假设。我们假设所提供的销售流水明细、批发价格、和损耗率数据是准确和可靠的，并且没有严重的错误或缺失。

2. 稳定性假设：我们假设销售、价格和损耗率在短期内保持相对稳定，不会受到外部因素（如季节性变化、市场突发事件等）的剧烈波动。

3. 损耗率均匀分布假设：我们可以假设损耗率在一定时间内是均匀分布的，不受外部因素的明显影响。

4. 市场竞争和替代品假设：我们可以忽略市场中其他竞争商超的影响，以及蔬菜的替代品对需求的潜在影响，以简化模型。

5. 最优解唯一性假设：我们可以假设在给定的条件下，存在一个唯一的最优解，即存在一种最佳的补货计划和定价策略，以最大化商超的收益。

# 三、 变量分析

为了更清晰地阐述我们所构建的相关模型，我们给出所用到模型变量及其相关含义。

表 1 变量解释

| 模型变量 | 变量含义 |
| --- | --- |
| $veg\_category_i, i=1,2,...,6$ | 6 种不同的蔬菜品类 |
| $veg\_item\_j, j=1,2,...,251$ | 251 种不同的蔬菜单品 |
| $sale_{j,d}, d=1,2,..,1085$ | 第 $d$ 天，第 $j$ 种蔬菜单品的销售价格 |
| $wholesale_{j,d}, d=1,2,..,1085$ | 第 $d$ 天，第 $j$ 种蔬菜单品的批发价格 |
| $sale\_num_{j,d}, d=1,2,...,1085$ | 第 $d$ 天，第 $j$ 种蔬菜单品的销售量 |
| $return_{j,d}$ | 第 $d$ 天，第 $j$ 种蔬菜单品的退货量 |
| $attrition_i, i=1,2,..,6$ | 第 $i$ 种蔬菜单品的损耗率 |
| $profit_d$ | 第 $d$ 天，商超的利润 |
| $supply_{j,d}$ | 第 $d$ 天，第 $j$ 种蔬菜单品的补货量 |
| $supply_{i,d}$ | 第 $d$ 天，第 $i$ 种蔬菜品类的补货量 |

# 四、 模型建立与求解

在本节，我们基于所提供的数据，综合运用数据分析与可视化技术、时序预测技术、非线性规划模型等，对所要解决的问题构建相应的数学模型并进行求解。

## 4.1 蔬菜各品类及单品销售量的分布规律及相互关系分析

首先，我们对蔬菜各品类的销售量分布规律及相互关系进行分析，结合（1）和（2）所提供的数据，我们可以得到蔬菜各品类的销售情况。图1展示了过去时间段内蔬菜各品类的销售量分布情况。

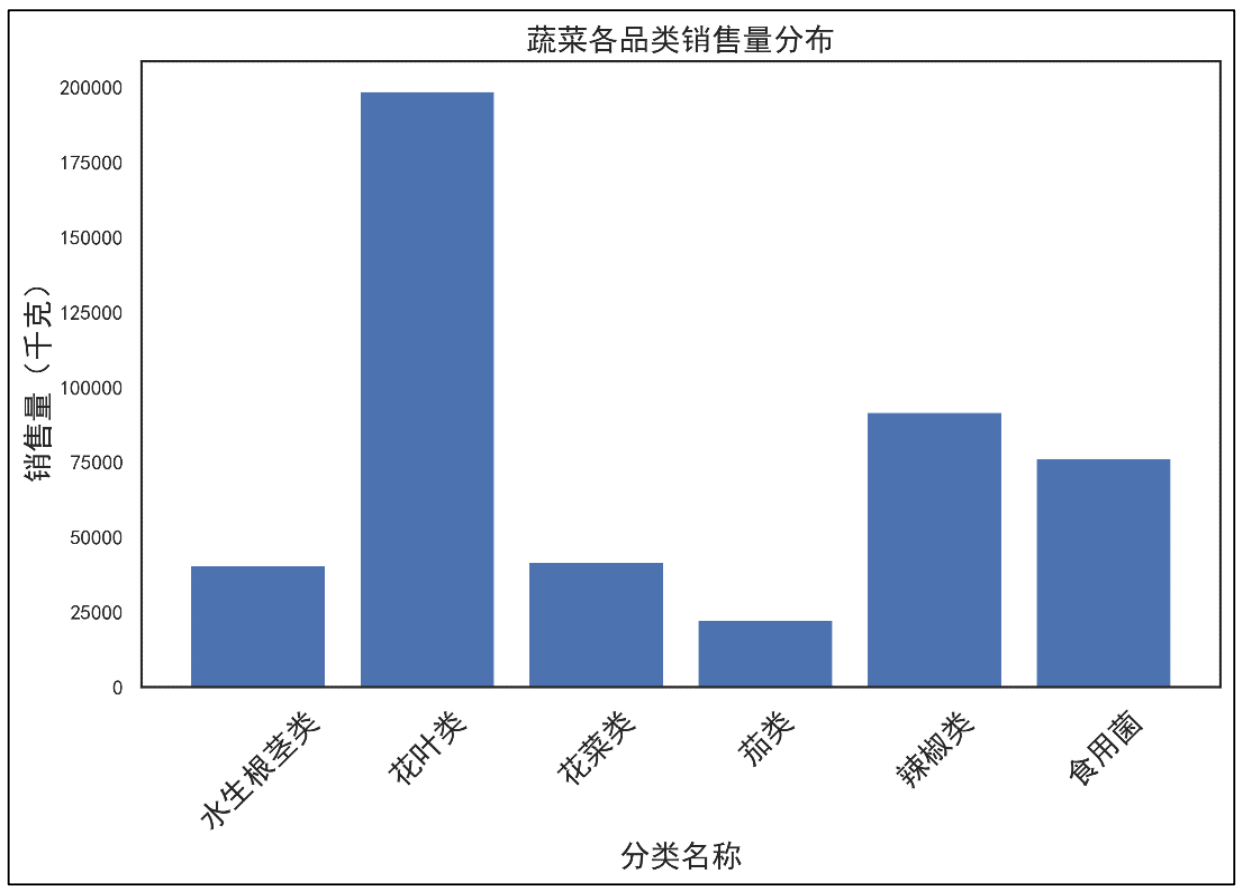

图 1 蔬菜各品类销售量分布情况

可以看到，蔬菜各品类中，花叶类销售量最大，其次是辣椒类、食用菌，水生根

茎类和花叶类的销售量近似，而茄类的销售量最低。

进一步地，我们以季度为周期，计算各蔬菜品类在各个季度的销售量，图2展示了从2020-09到2023-05期间内，9个季度蔬菜品类的销售量变化。

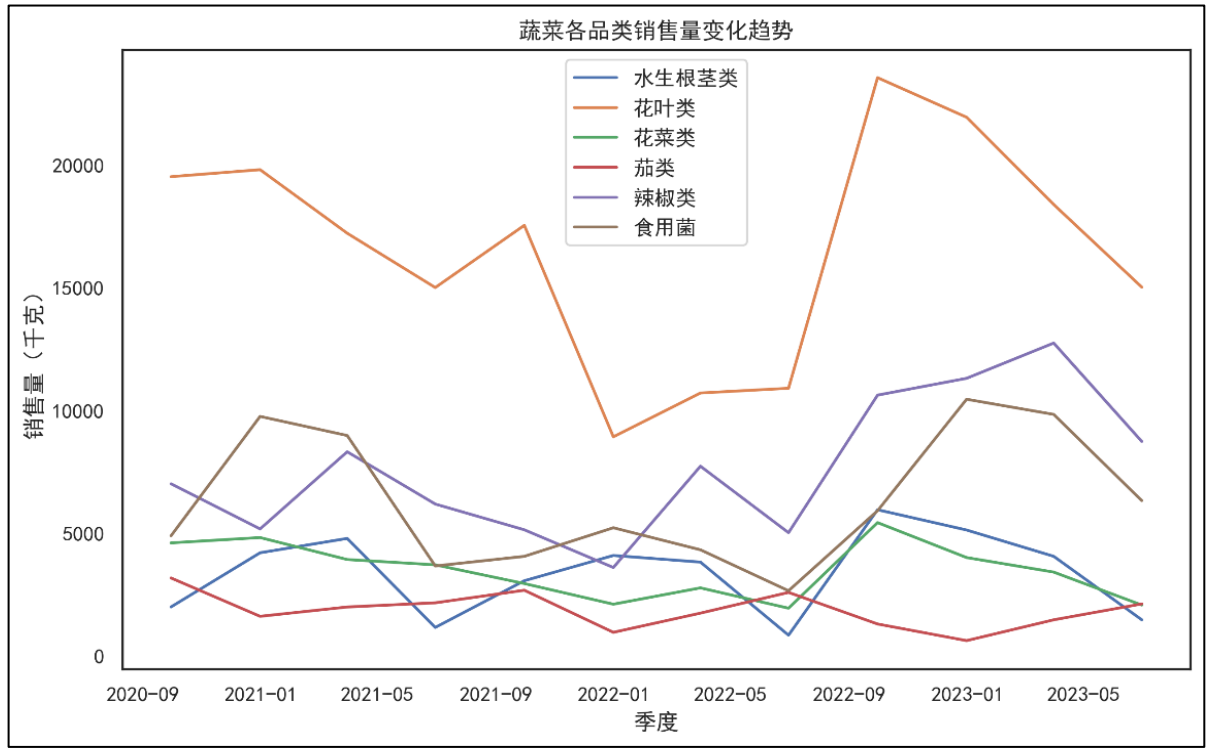

图 2 各蔬菜品类销售量季度变化情况

可以看到，蔬菜品类的销售量受到时间影响，以茄类为例，其销量在一季度往往会有下降的趋势，花叶类蔬菜的销售高峰往往出现在秋冬季，而辣椒类往往出现在春夏季度，所以商超在制定补货和定价策略时，应该考虑季度因素。

同样以季度为周期，图 3 展示了蔬菜各品类的平均销售单价和销售总价的变化趋势，蔬菜各品类的平均销售单价计算公式如下：

$$sale\_avg_i = \frac{\sum_d \sum_j sale_{j,d} \times sale\_num_{j,d}}{\sum_d \sum_j sale\_num_{j,d}}$$

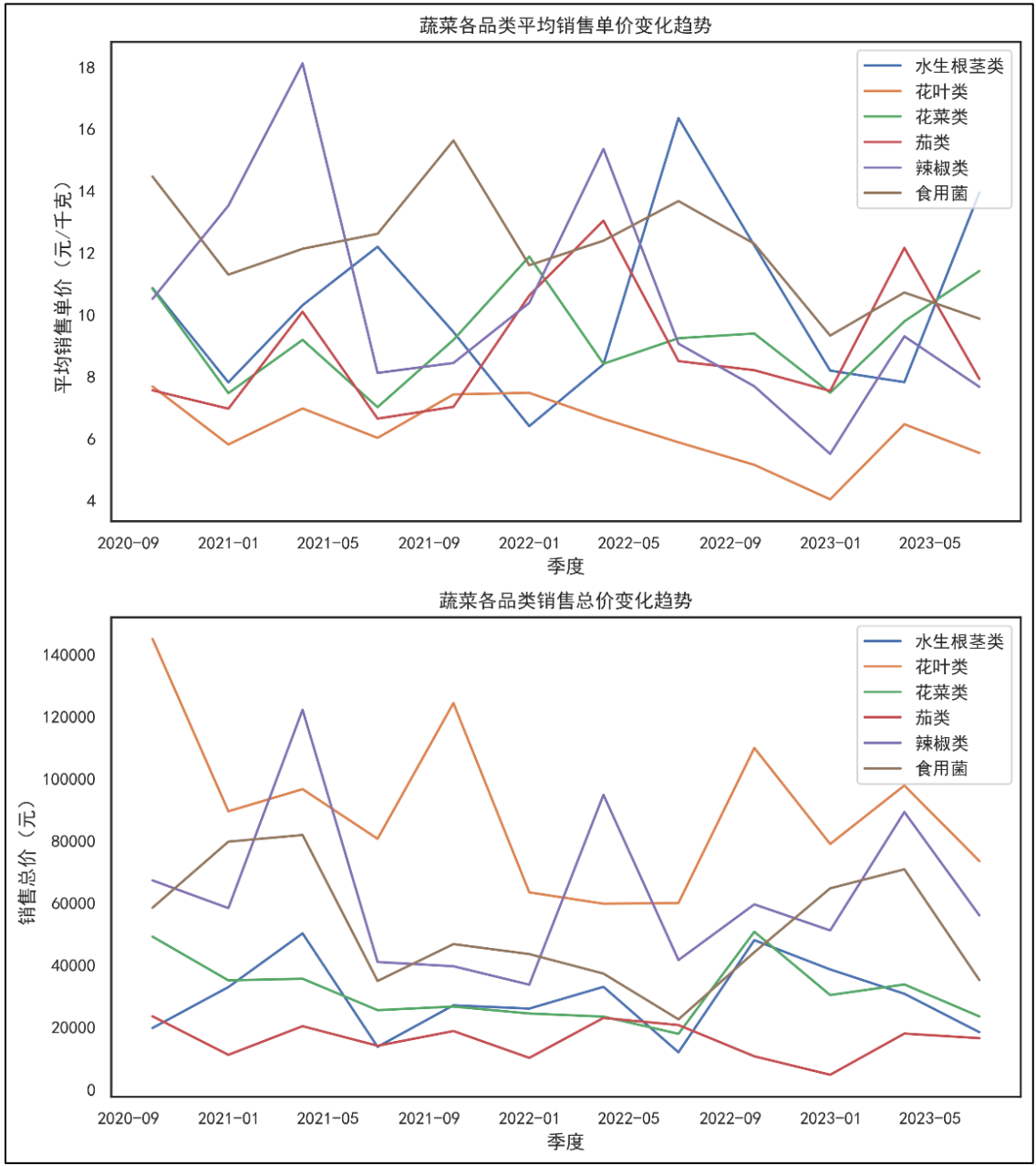

图 3 蔬菜各品类销售单价及总价随季度变化趋势

由图中可以看到，花叶类蔬菜的销售单价在每个季度近乎都是最低的，但是其销售总价非常高，结合图 2 的销售量情况，说明销售单价较低的蔬菜品类具有较高的销售量，销售量和销售定价之间存在着一定的关系。

结合（1）、（2）、（3）和（4）所提供的数据，我们可以计算蔬菜各品类的利润变化趋势，蔬菜各品类的利润计算公式如下所示：

$$profit_i = \sum_d \sum_j sale_{j,d} \times sale\_num_{j,d} - \sum_d \sum_j sale_{j,d} \times return_{j,d}$$
$$- \sum_d \sum_j wholesale_{j,d} \times (1 + attrition_i) \times sale\_num_{j,d}$$

图 4 展示了以季度为周期，蔬菜各品类的利润变化趋势。

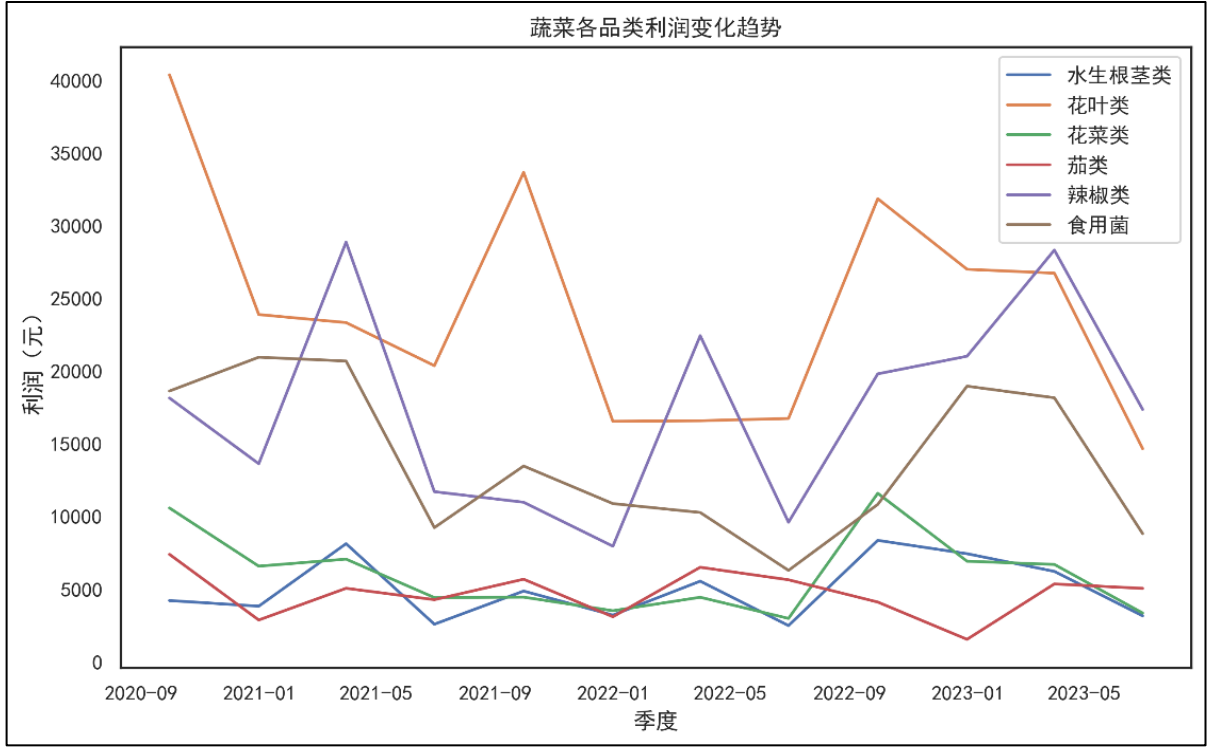

图 4  蔬菜各品类利润随季度变化趋势

可以看到，蔬菜各品类利润变化与销售总价的变化大致相同，花叶类蔬菜的售卖由于销售量大，给商超带来的利润也是最多的，其次是辣椒类和食用菌类，茄类，花菜类和水生根茎类的利润大致相同。

以季度为周期，我们利用皮尔逊相关系数计算蔬菜各品类销售量之间的关联关系，皮尔逊相关系数的计算方式如下：

$$\rho_{X,Y} = \frac{\sum_{i=1}^{n}(X_i - \bar{X})(Y_i - \bar{Y})}{\sqrt{\sum_{i=1}^{n}(X_i - \bar{X})^2}\sqrt{\sum_{i=1}^{n}(Y_i - \bar{Y})^2}}$$

其中，$X_i$ 和 $Y_i$ 是两个变量 $X$ 和 $Y$ 的观测值，$\bar{X}$ 和 $\bar{Y}$ 分别是 $X$ 和 $Y$ 的均值，$n$ 是观测值的数量。

图 5 展示了蔬菜各品类销售量相关系数矩阵的热力图。

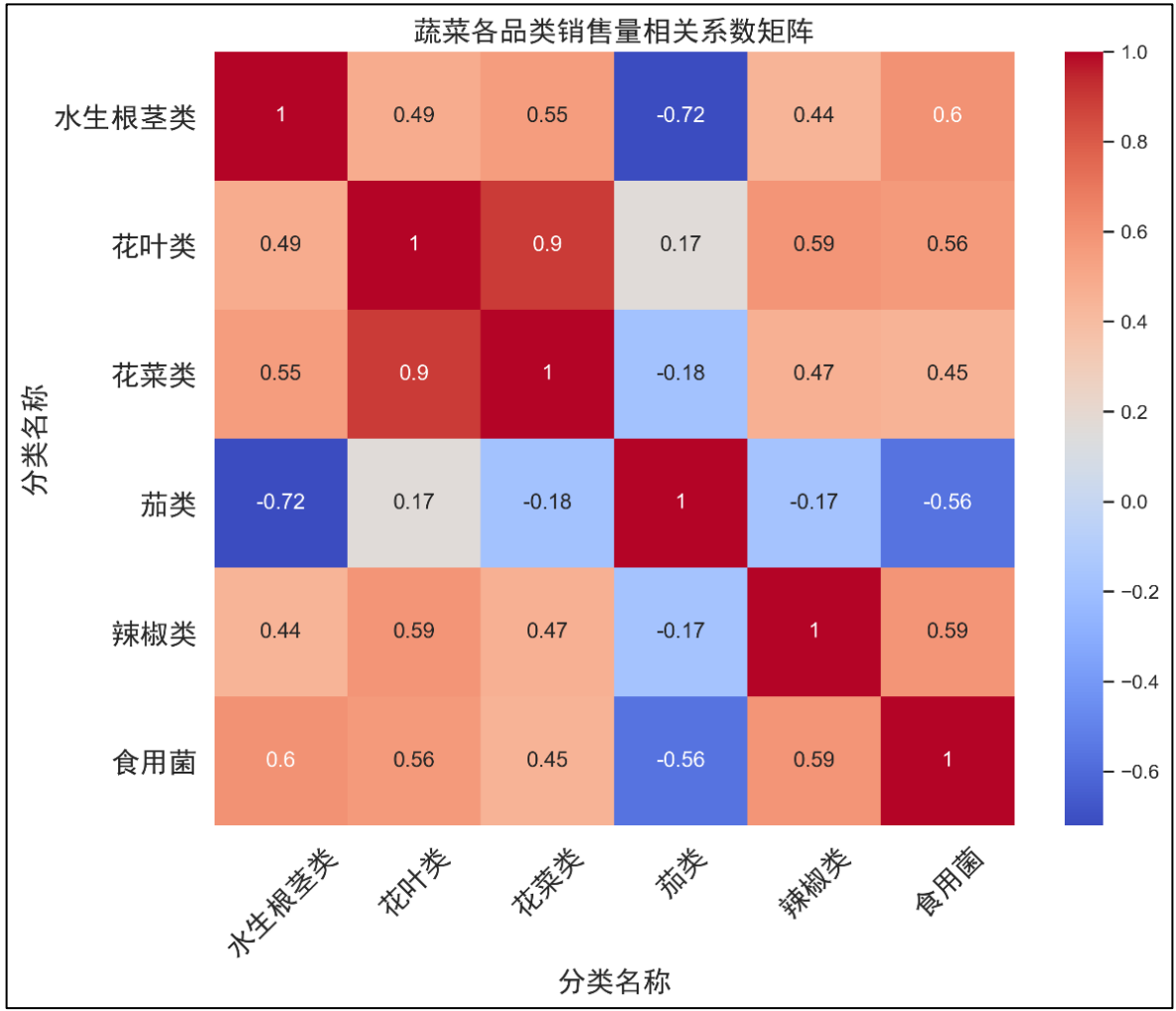

图 5  蔬菜各品类相关系数矩阵热力图

这个热力图展示了蔬菜各品类销售量之间的相关系数矩阵。相关系数矩阵是一个方阵，其中每个元素是两个变量之间的相关系数。在这个热力图中，每个单元格的颜色表示对应品类之间的相关系数大小，颜色越深表示相关性越强，颜色越浅表示相关性越弱。根据相关系数矩阵，我们可以看到以下几种蔬菜品类之间的相关性较强：

- 花叶类和花菜类的相关系数为 0.895994，相关性较强。
- 水生根茎类和食用菌的相关系数为 0.597366，相关性较强。
- 花叶类和水生根茎类的相关系数为 0.488618，相关性较强。

另外，我们也可以看到茄类和水生根茎类之间的相关系数为-0.719960，呈现出负相关性，即当水生根茎类的产量增加时，茄类的产量反而会下降。

相似地，我们进一步分析蔬菜单品销售量的分布情况和关联情况。图 6 展示了各蔬菜单品销售量的分布情况。

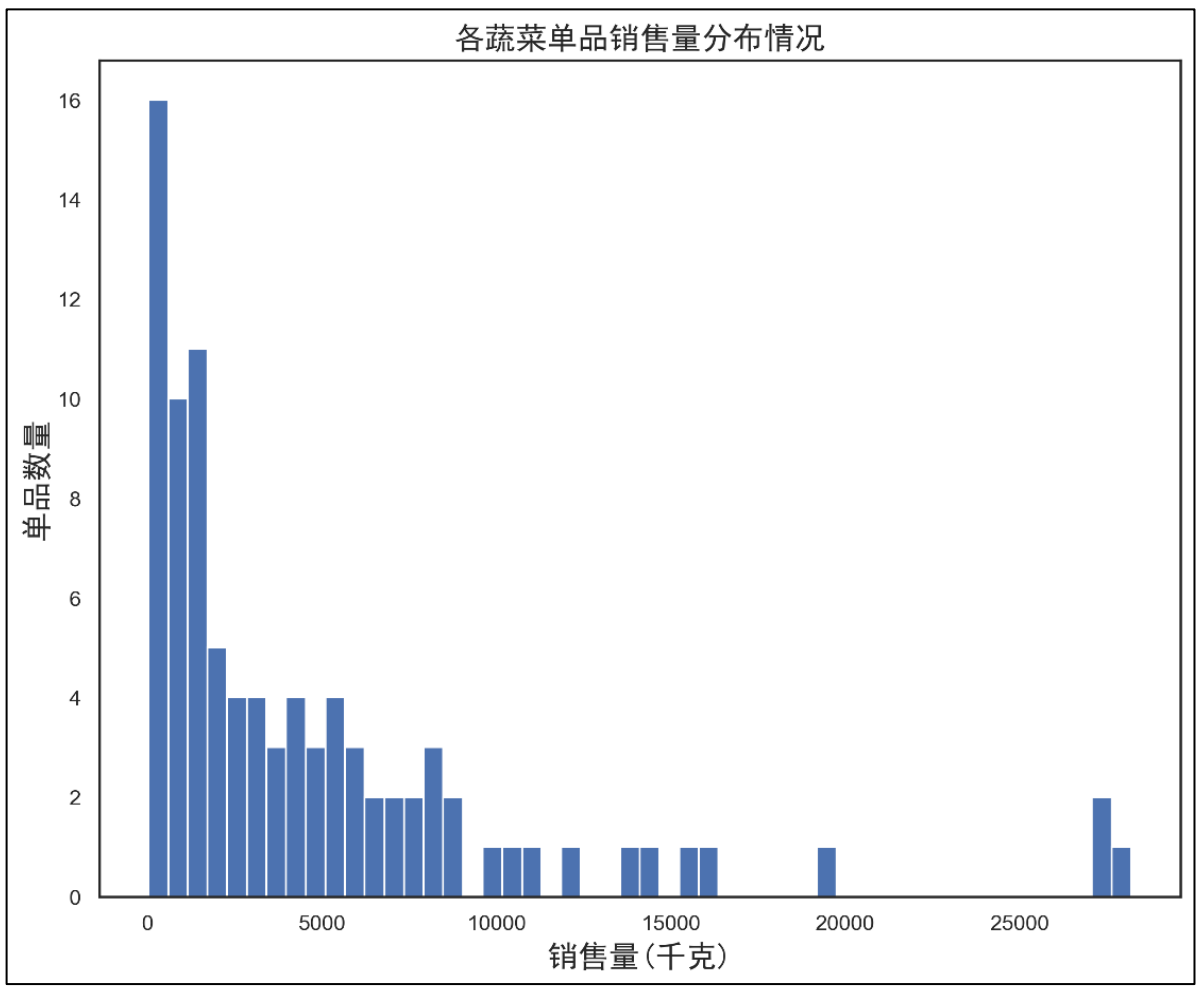

图 6 蔬菜单品销售量分布情况

这个直方图展示了各蔬菜单品销售量的分布情况。横轴表示销售量（单位为千克），纵轴表示单品数量。直方图将销售量分成了 50 个区间，每个区间的高度表示该区间内单品数量的数量。可以看出，销售量在 0-5000 千克之间的单品数量最多，而销售量在 5000-15000 千克之间的单品数量逐渐减少。同时，销售量在 15000 千克以上的单品数量非常少。

图 7 展示了销售量最高的前 10 个蔬菜单品的销售量。我们可以看到销售量最高的前十个单品名称及其销售量。其中，销售量最高的单品名称为芜湖青椒（1），销售量为 28199.151。而销售量最低的单品名称为小米椒（份），销售量为 10861.0。销售量最高的前三个单品分别为芜湖青椒（1）、西兰花和净藕（1），说明这三个单品在市场上的需求最大。

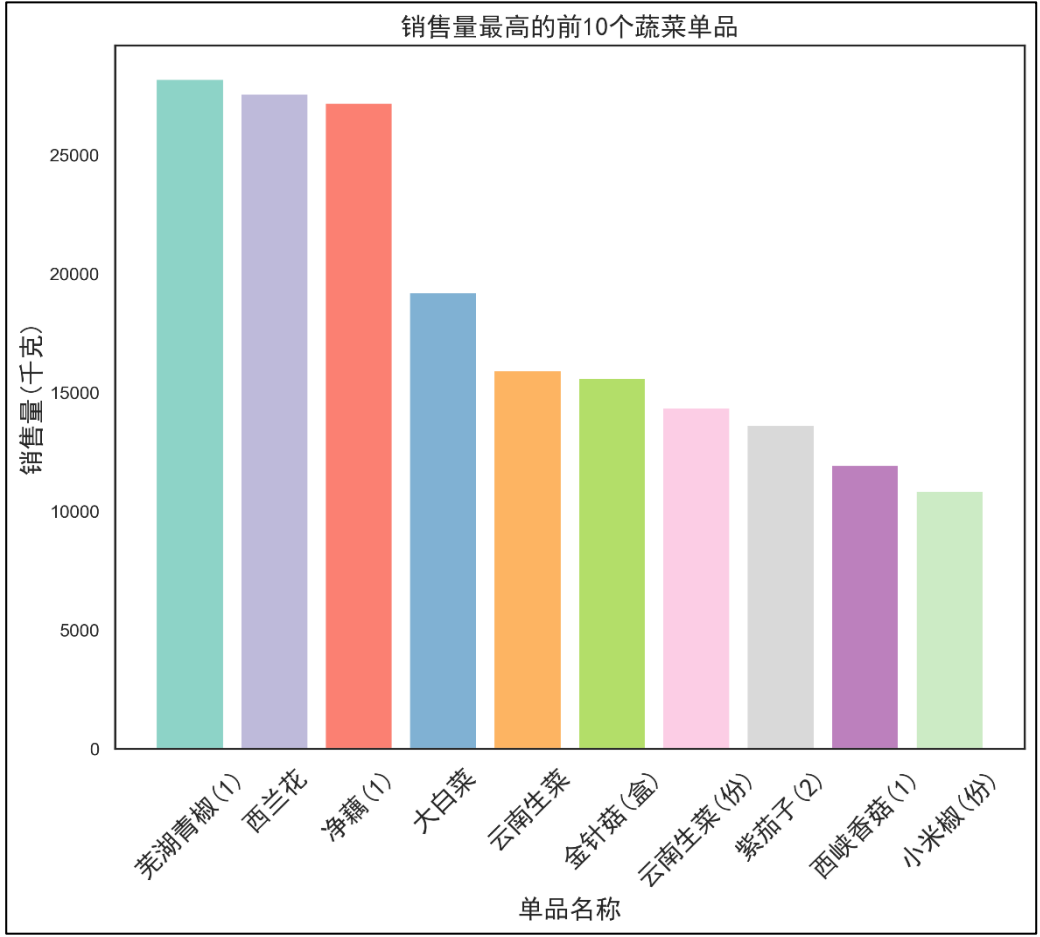

图 7 销售量最高的前 10 个蔬菜单品

图 8 展示了以季度为周期蔬菜销售量前 10 的单品的变化趋势。

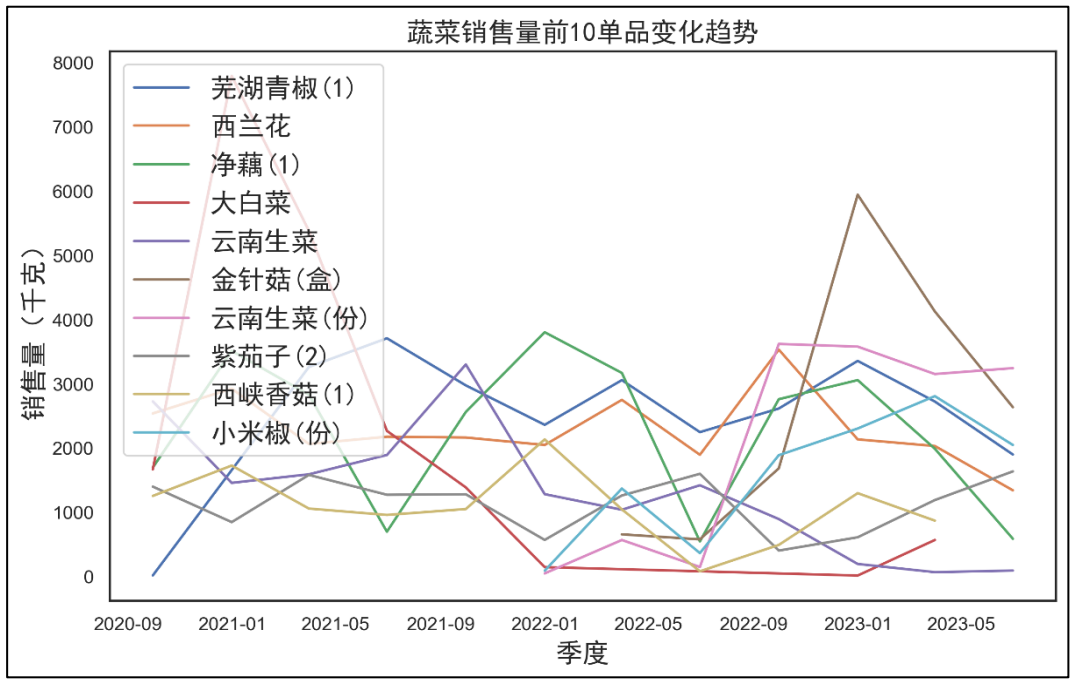

图 8 蔬菜销售量前 10 单品随季度变化图示

同样，我们计算了不同蔬菜单品的销售量的相关系数矩阵以分析其关联关系，图 9 展示了销售量前 15 的蔬菜单品销售量相关系数矩阵热力图。

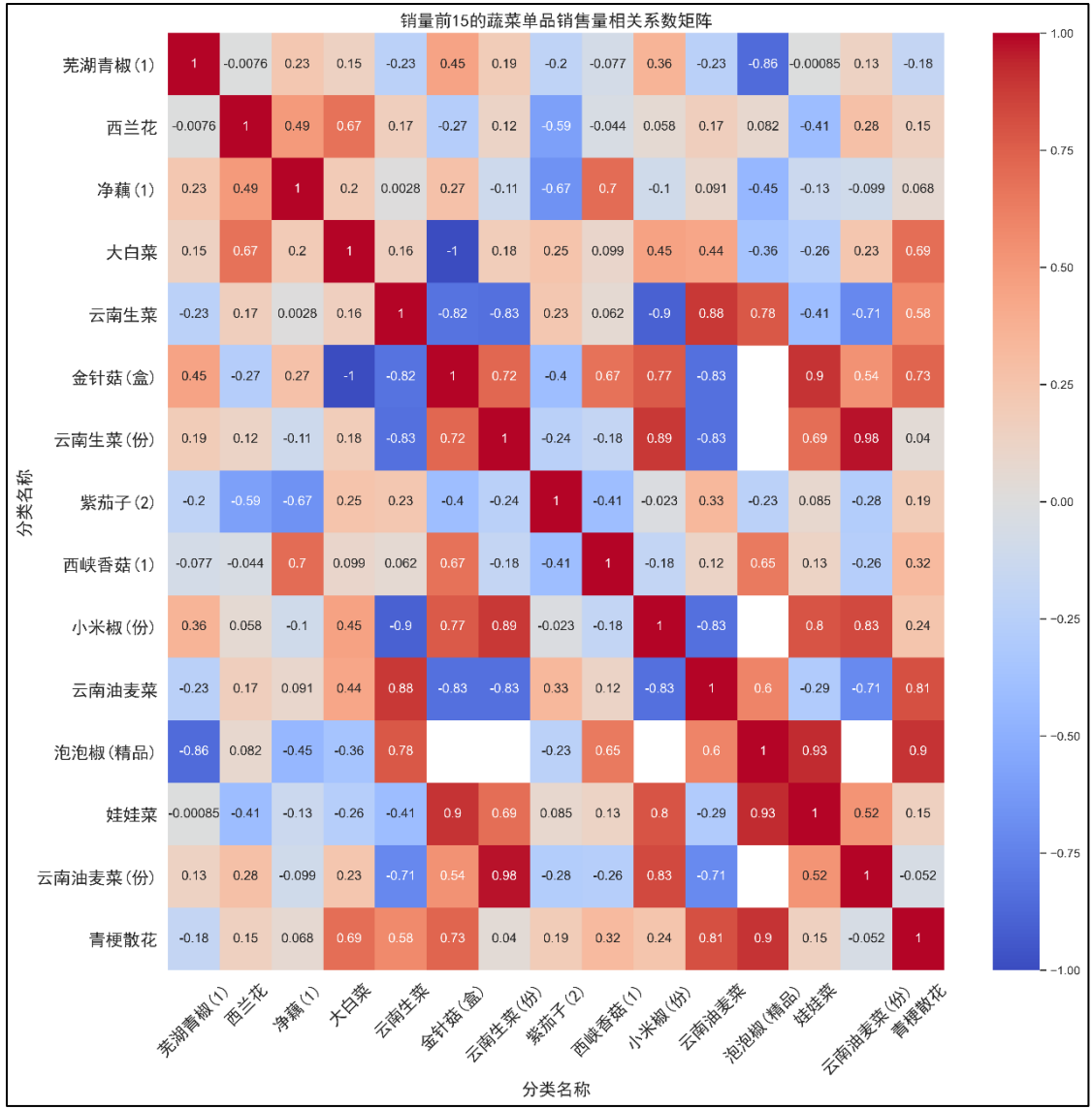

图 9 销量前 15 的蔬菜单品销售量热力图

根据相关系数矩阵，我们可以看到以下几种蔬菜单品之间的相关性较强：

● 云南生菜和云南油麦菜之间的相关系数为 0.883445，相关性较强。

● 云南生菜和金针菇（盒）之间的相关系数为 0.724516，相关性较强。

● 云南生菜和云南生菜（份）之间的相关系数为-0.831510，呈现出负相关性，即当云南生菜的销售量增加时，云南生菜（份）的销售量反而会下降。

● 云南生菜和泡泡椒（精品）之间的相关系数为 0.778169，相关性较强。

另外，我们也可以看到大白菜和金针菇（盒）之间的相关系数为-1.000000，呈现出负相关性，即当大白菜销售量增加时，金针菇（盒）的销售量反而会下降。

## 4.2 蔬菜品类未来一周的日补货总量和定价策略

为了制定商超各蔬菜品类未来一周的日补货总量和定价策略，使得商超收益最大化，我们首先需要分析各蔬菜品类销售总量和成本加成定价的关系，以确定销售定价对销售总量的影响。由于以蔬菜品类为单位进行分析，我们首先计算蔬菜品类每天的平均销售定价，以该定价代替成本加成定价，该定价计算公式如下：

$$sale\_avg_{i,d} = \frac{\sum_{j \in i} sale_{j,d} \times sale\_num_{j,d}}{\sum_{j \in i} sale\_num_{j,d}}$$

各蔬菜品类每天的销售总量计算公式如下：

$$sale\_num_{i,d} = \sum_{j \in i} sale\_num_{j,d}$$

我们考虑不同的函数用于拟合不同的蔬菜品类平均销售定价与销售总量的关系，包括线性函数、幂函数和对数函数，并选择拟合效果最佳的模型作为定价-销量模型 *func*。图 10 到 16 展示了不同函数对六种蔬菜品种的拟合效果。最终选择的模型参数如下表所示：

表 2 拟合模型参数表

| 分类名称 | 模型名称 | 模型参数 |
| --- | --- | --- |
| 水生根茎类 | 对数函数 | [-26.11035516, 2.51727164, 83.60511972] |
| 花叶类 | 对数函数 | [-36.00775817, 2.35435122, 209.756959] |
| 花菜类 | 线性函数 | [-2.89165146, 64.2021206] |
| 茄类 | 对数函数 | [-8.16177241, 1.62367199, 36.4701182] |
| 辣椒类 | 对数函数 | [-17.07560033, 3.10816607, 105.16379865] |
| 食用菌 | 线性函数 | [-3.28878025, 92.53732638] |

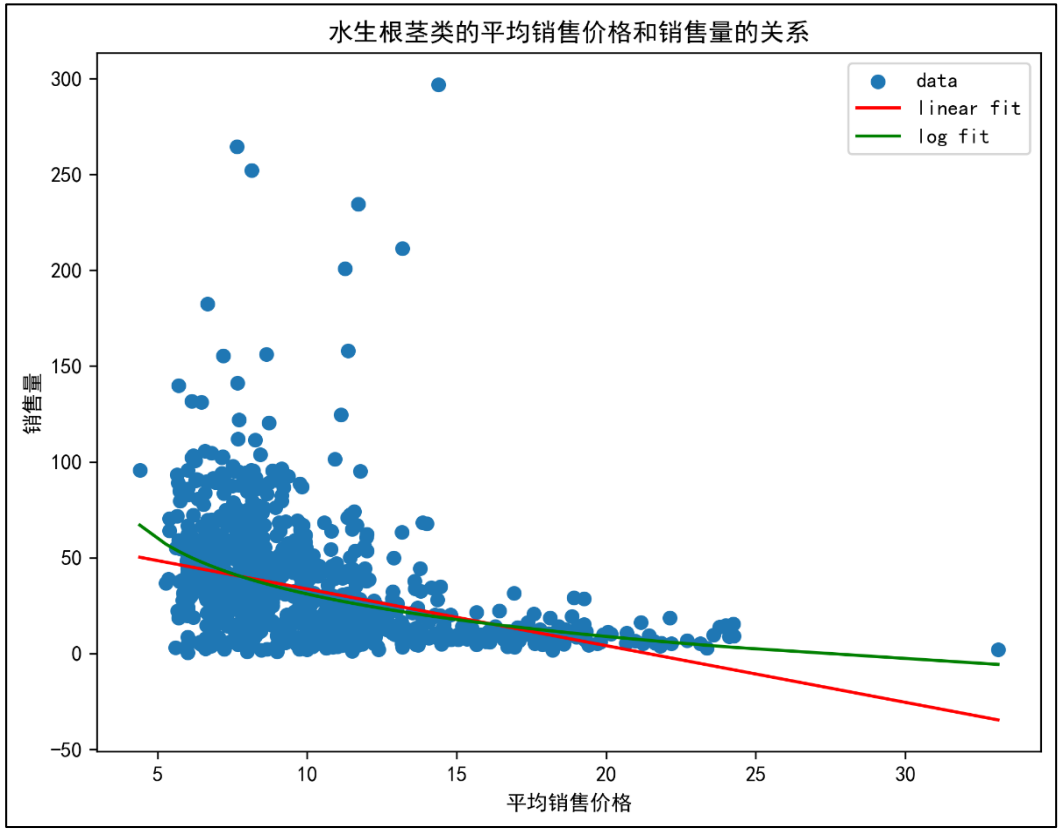

图 10 水生根茎类平均销售价格和销售量关系拟合图

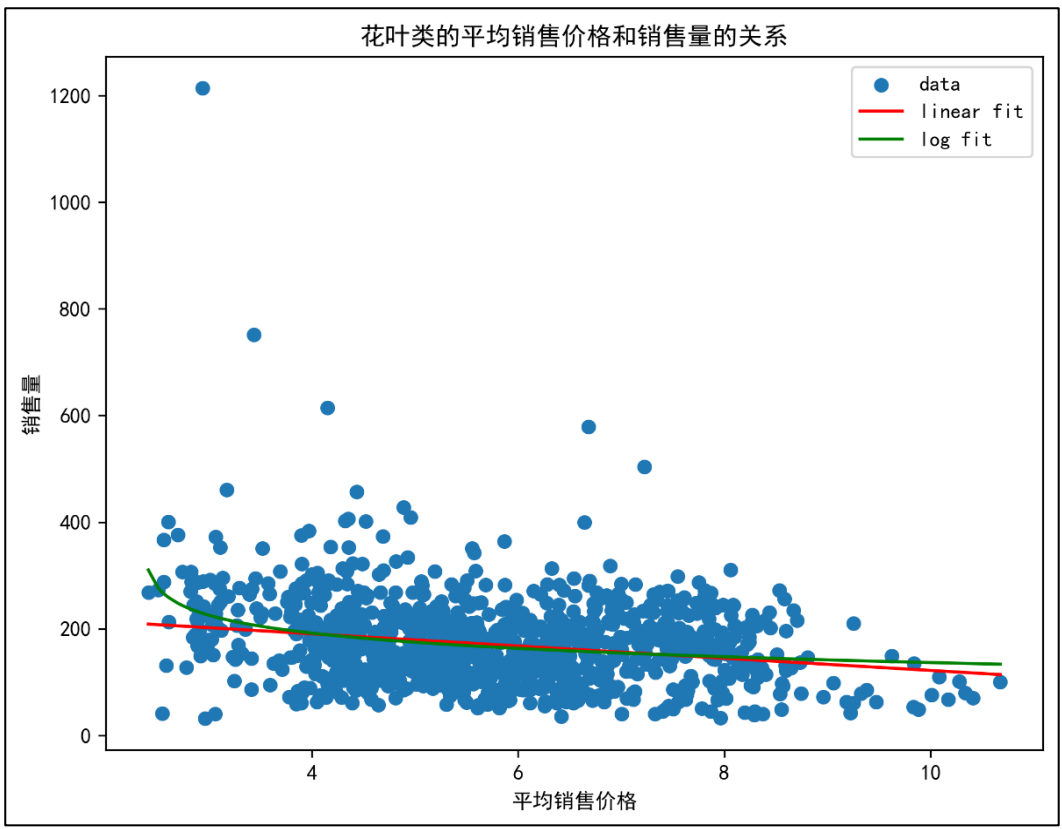

图 11 花叶类平均销售价格和销售量关系拟合图

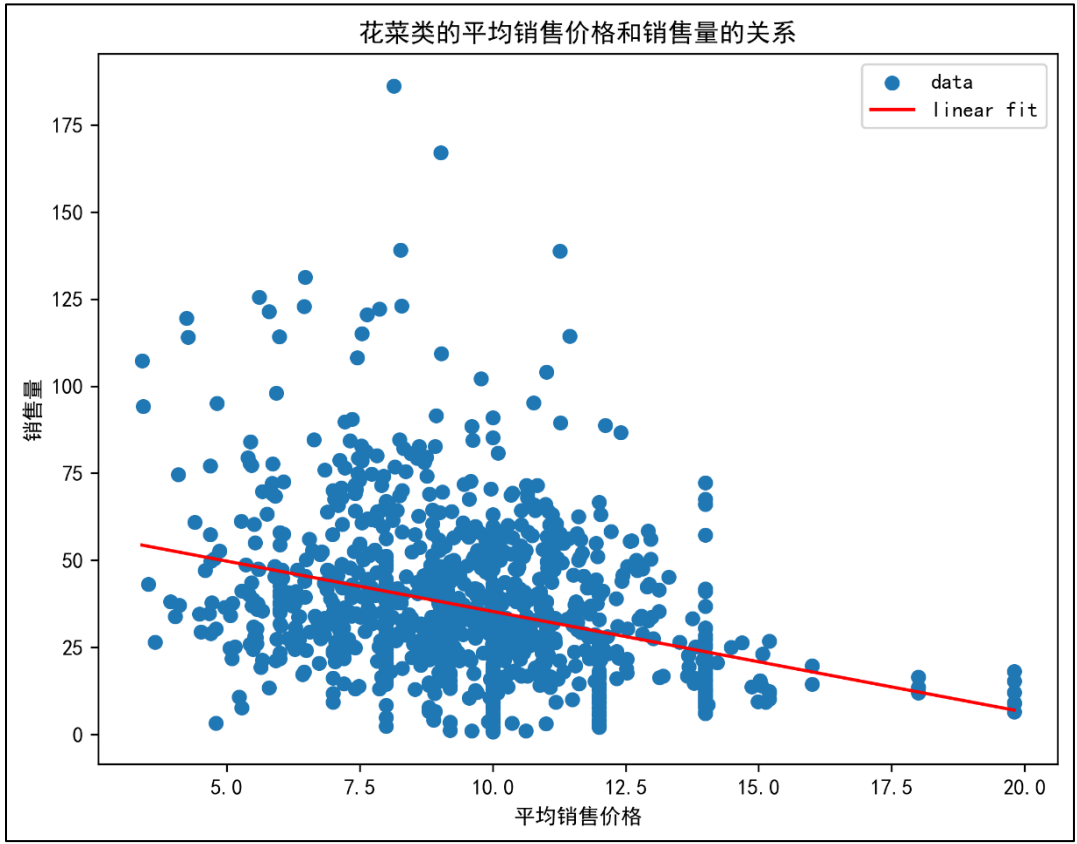

图 12 花菜类平均销售价格和销售量关系拟合图

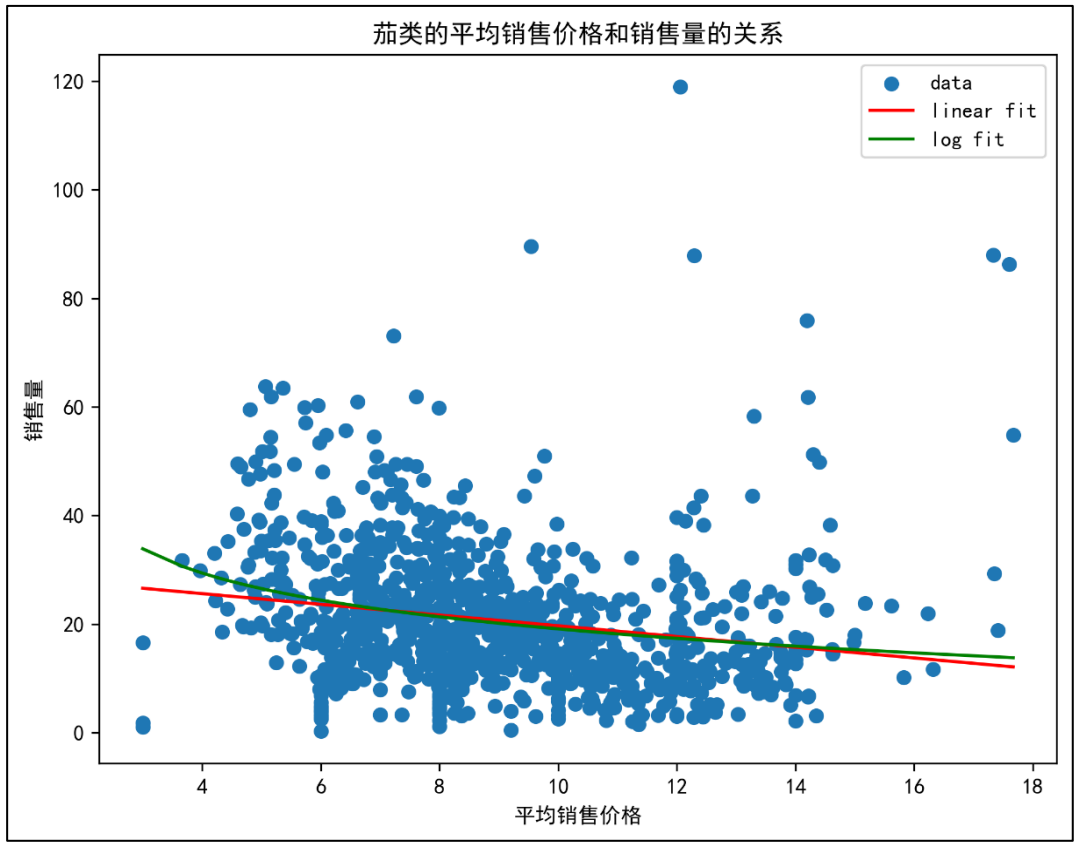

图 13 茄类平均销售价格和销售量关系拟合图

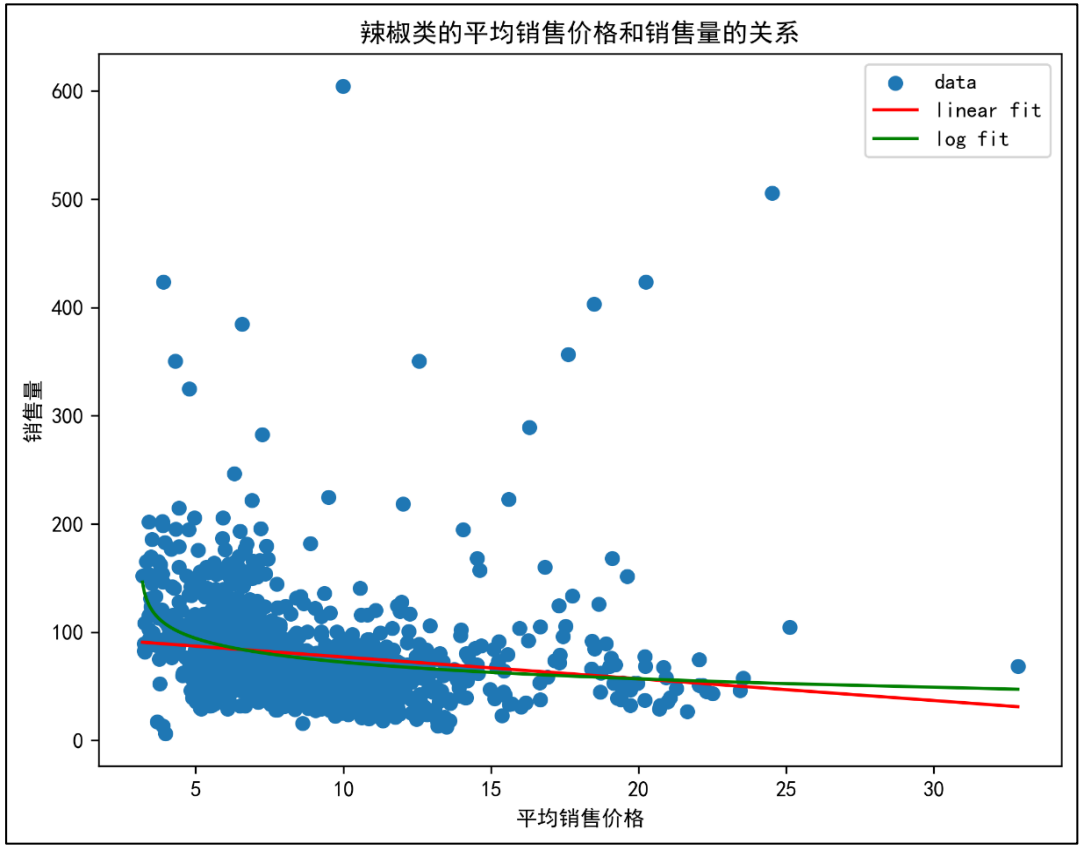

图 14 辣椒类平均销售价格和销售量关系拟合图

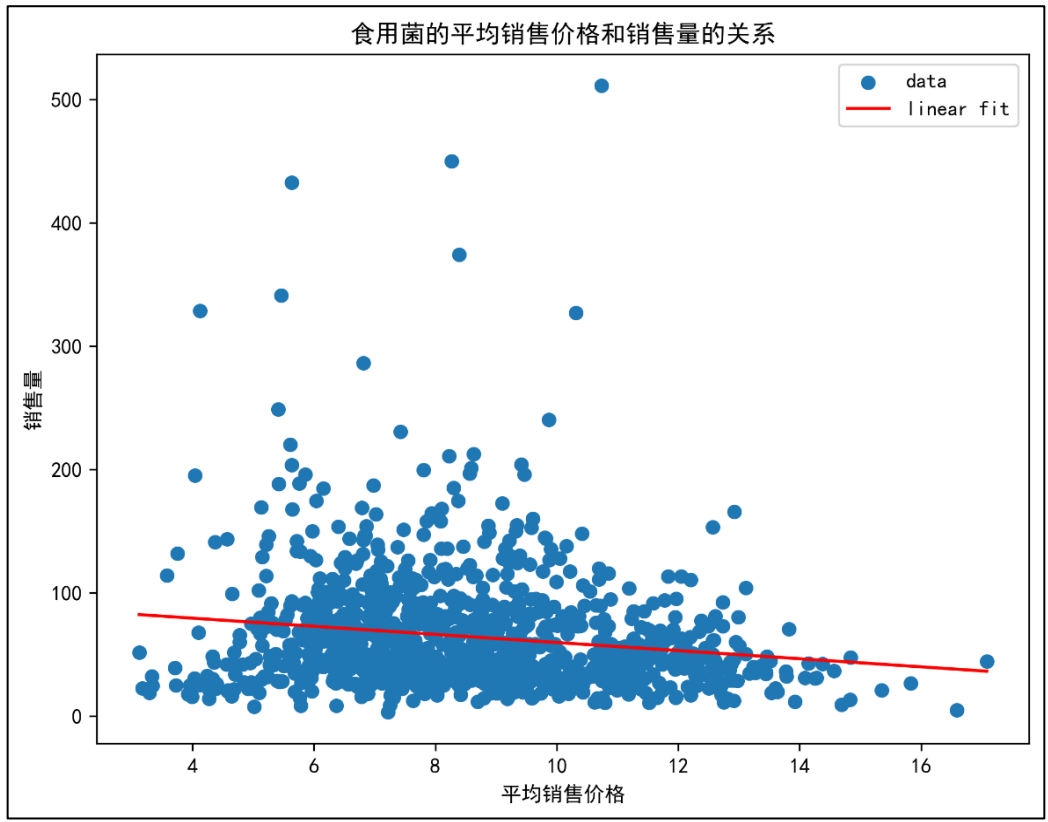

图 15 食用菌类平均销售价格和销售量关系拟合图

为了获取商品利润，我们还需要对蔬菜品类的平均批发价进行预测，蔬菜品类每天的平均批发价计算公式如下：

$$whole\_avg\_sale_{i,d} = \frac{\sum_{j \in i}(1+attrition_i) \times sale\_num_{j,d} \times wholesale_{j,d}}{\sum_{j \in i}(1+attrition_i) \times sale\_num_{j,d}}$$

图 16 展示了以月为周期，各蔬菜品类平均批发价格的变化。

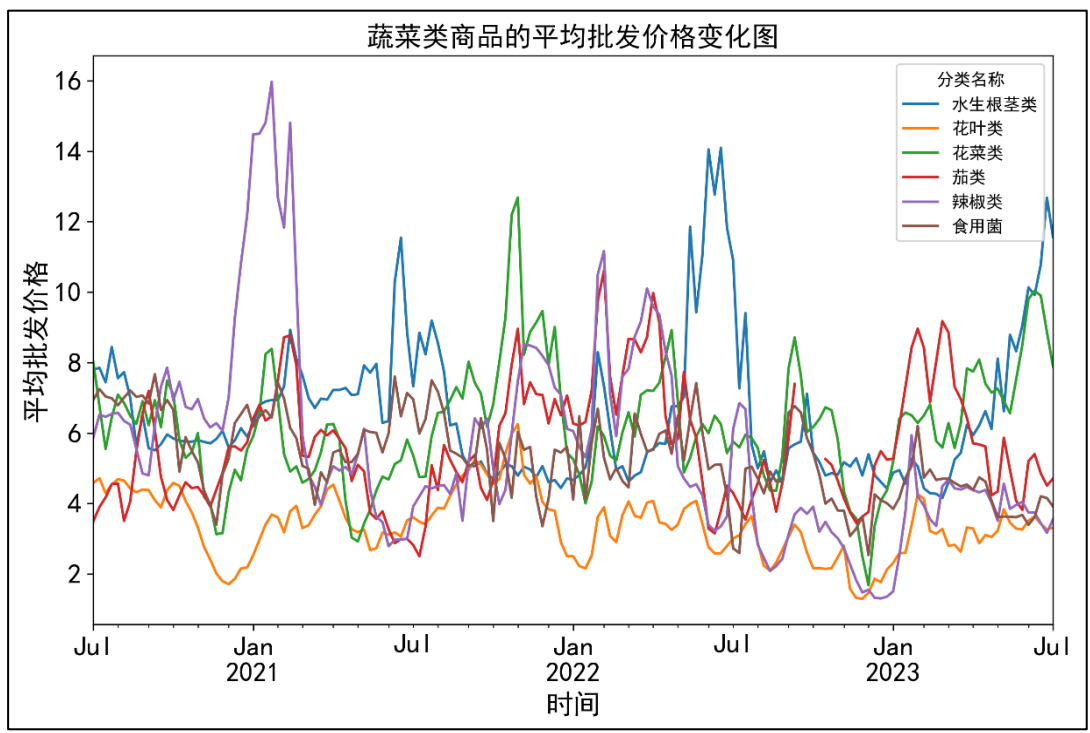

图 16 蔬菜类商品的平均批发价格变化

我们利用 ARIMA 模型对蔬菜品类未来 7 天的平均批发价进行预测。

自回归差分移动平均模型（ARIMA），是一种时间序列分析中常用的统计模型，用于描述和预测时间序列数据的变化趋势。ARIMA 模型基于以下核心思想：时间序列数据中的趋势和周期性可以通过对数据进行自回归、差分和移动平均的操作来捕捉和建模。ARIMA 模型的名称反映了其主要组成部分：

自回归（AR，Autoregressive）：这部分考虑了时间序列数据中过去时间点的值对当前值的影响。具体来说，它表示当前时间点的值与前一时间点、前两个时间点等的线性组合。

差分（I，Integrated）：差分是指对时间序列数据进行减法操作，目的是消除数据

中的趋势或季节性。通过差分操作，原始的非平稳时间序列可以转化为平稳的序列。

移动平均（MA，Moving Average）：这部分考虑了时间序列数据中过去时间点的误差对当前值的影响。具体来说，它表示当前时间点的误差与前一时间点、前两个时间点等的线性组合。

ARIMA 模型的参数通常表示为 p、d 和 q，分别对应了自回归、差分和移动平均的阶数。选择合适的 p、d 和 q 值是建立 ARIMA 模型的关键步骤。

我们利用模型的赤池信息准则（Akaike Information Criterion，AIC）选择最佳的模型参数为（2,2,2），对未来七天的蔬菜品类平均批发价格进行了预测。

得到蔬菜品类平均批发价格和定价-销售量模型之后，我们构建了非线性规划模型，以求解最佳的补货总量和定价策略。所构建的非线性规划模型如下：

$$\text{objetive:} \quad \max \ profit_{i,d} = sale_{i,d} \times sale\_num_{i,d} - supply_{i,d} \times wholesale_{i,d}$$

$$\text{s.t.:} \quad supply_{i,d} > sale\_num_{i,d} + attrition_i \times supply_{i,d} \quad (1)$$

$$sale_{i,d} > wholesale_{i,d} \quad (2)$$

$$sale_{i,d}, supply_{i,d} > 0 \quad (3)$$

$$sale\_num_{i,d} = func(sale_{i,d}) \quad (4)$$

其中，目标函数为最大化商超第 $d$ 天在蔬菜品类 $i$ 上的利益，含义为销售总价-批发总价，$supply_{i,d}$ 为商品 $i$ 在 $d$ 天的补货量，$sale_{i,d}$ 为商品 $i$ 在 $d$ 天的销售量；约束函数 1 为补货量大于销售量加损耗量，约束函数 2 为销售价大于成本批发价，约束函数 3 为补货量和销售定价大于 0。约束函数 4 为定价-销售量函数，通过该函数可以由销售定价确定销售量，带入数据，求解得到商超未来 7 天的蔬菜品类日补货量和定价策略如下表所示：

表 3  商超未来 7 天的蔬菜品类日补货量和定价策略

| 日期 | 类别 | 进货量（kg） | 定价（元/kg） |
| --- | --- | --- | --- |
| 1 | 水生根茎类 | 10.95334 | 19.62878 |
| 2 | 水生根茎类 | 10.94375 | 19.6342 |
| 3 | 水生根茎类 | 10.82813 | 19.69978 |
| 4 | 水生根茎类 | 10.80803 | 19.71121 |
| 5 | 水生根茎类 | 10.79247 | 19.72005 |

| 日期 | 类别 | 进货量（kg） | 定价（元/kg） |
| --- | --- | --- | --- |
| 6 | 水生根茎类 | 10.78517 | 19.72421 |
| 7 | 水生根茎类 | 10.77935 | 19.72752 |
| 1 | 花叶类 | 159.7414 | 9.440909 |
| 2 | 花叶类 | 159.9985 | 9.396931 |
| 3 | 花叶类 | 160.0699 | 9.384782 |
| 4 | 花叶类 | 159.9614 | 9.403269 |
| 5 | 花叶类 | 160.1486 | 9.371391 |
| 6 | 花叶类 | 160.0068 | 9.395525 |
| 7 | 花叶类 | 160.2002 | 9.362627 |
| 1 | 花菜类 | 22.07802 | 15.75169 |
| 2 | 花菜类 | 22.1002 | 15.74521 |
| 3 | 花菜类 | 22.08577 | 15.74943 |
| 4 | 花菜类 | 22.09992 | 15.74529 |
| 5 | 花菜类 | 22.08922 | 15.74842 |
| 6 | 花菜类 | 22.10044 | 15.74514 |
| 7 | 花菜类 | 22.09229 | 15.74752 |
| 1 | 茄类 | 17.07341 | 14.00598 |
| 2 | 茄类 | 17.08081 | 13.99551 |
| 3 | 茄类 | 17.06499 | 14.0179 |
| 4 | 茄类 | 17.06977 | 14.01114 |
| 5 | 茄类 | 17.0663 | 14.01606 |
| 6 | 茄类 | 17.06447 | 14.01864 |
| 7 | 茄类 | 17.06293 | 14.02083 |
| 1 | 辣椒类 | 76.95735 | 11.01945 |
| 2 | 辣椒类 | 76.7876 | 11.09116 |
| 3 | 辣椒类 | 76.96742 | 11.01522 |
| 4 | 辣椒类 | 76.81373 | 11.08008 |
| 5 | 辣椒类 | 76.98788 | 11.00663 |
| 6 | 辣椒类 | 76.8376 | 11.06997 |
| 7 | 辣椒类 | 77.00889 | 10.99781 |
| 1 | 食用菌 | 54.24866 | 13.20099 |
| 2 | 食用菌 | 52.53689 | 13.67229 |
| 3 | 食用菌 | 51.98256 | 13.82492 |
| 4 | 食用菌 | 53.4264 | 13.42738 |
| 5 | 食用菌 | 51.83244 | 13.86625 |
| 6 | 食用菌 | 53.38939 | 13.43758 |
| 7 | 食用菌 | 52.05502 | 13.80497 |

## 4.3 各蔬菜单品未来一天的日补货总量和定价策略

商超补货时需要控制可售单品总数，且满足单品订购量的最小陈列量要求，在尽量满足市场对各品类蔬菜商品需求的前提下，使得商超收益最大化。我们假设 7 月 1 日所能补货的单品与上一周的可售品种相同，图 17 展示了 2023 年 6 月 24 到 30 日蔬菜单品的销售情况，总共有 49 种可以采购的单品。

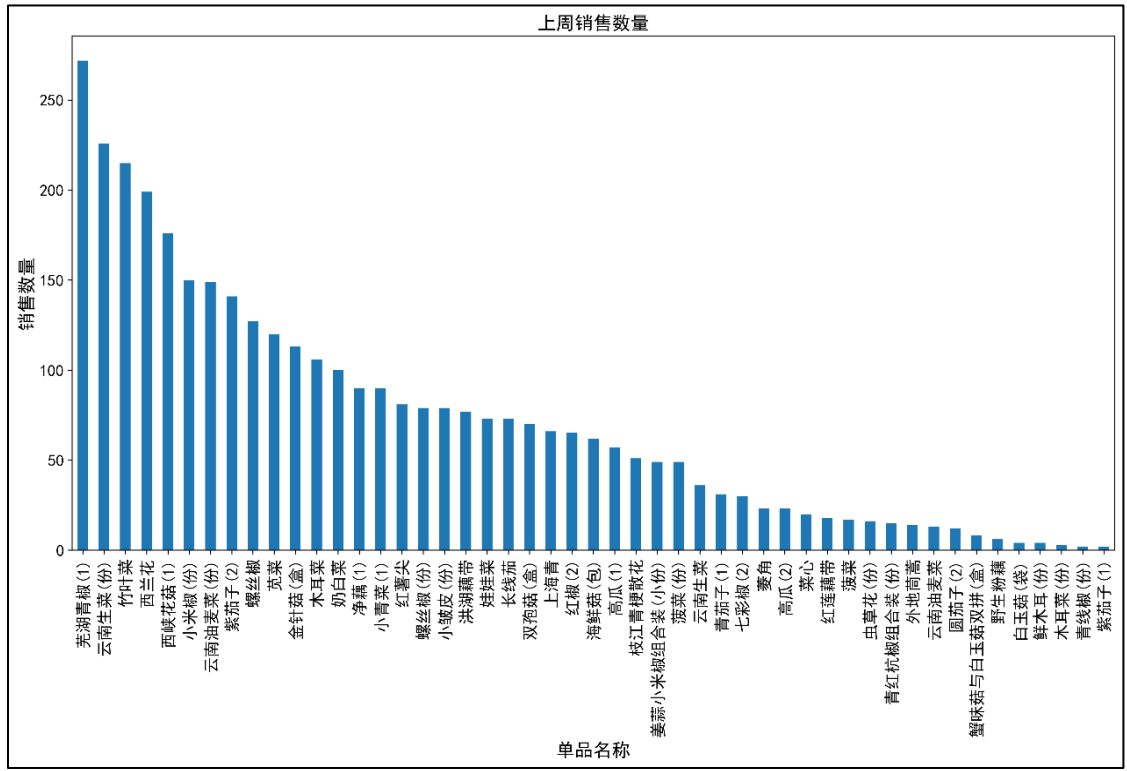

图 17 前一周蔬菜单品销售情况分布

与 4.2 类似，我们首先对各单品未来一天的批发价格进行了预测，在这里我们采用历史平均值预测方法得到未来的批发价格 $wholesale_{j,d}$，定义非线性规划模型如下：

objective: $\max \left( profit_d = select_d \times sale_d \times sale\_num_d - select_d \times supply_d \times wholesale_d \right)$

s.t.: $select_d \times supply_{j,d} > select_d \times sale\_num_{j,d} + select_d \times attrition_i \times supply_{j,d}, where\ j \in i$

$sale_{j,d} > wholesale_{j,d}$

$select_d \times supply_{j,d} > select_d \times 2.5$

$27 \leq sum(select_d) \leq 33$

$select_d \times supply_{j,d} = supply_{j,d}$

$$select_d \times sale_{j,d} = sale_{j,d}$$

$$sale\_num_{j,d} = func(sale_{j,d}), where\ j \in i$$

$$num(i) = 6$$

$$select_d = [random.randint(0,1)\ for\_in\ range(49)]$$

其中，目标函数 *objective* 为最大化商超的销售利润，为每个蔬菜单品的销售利润的总和，在这里我们使用了一个选择列表 $select_d$ 来控制蔬菜单品的采购数量，其为一个长度为 49，值为 0 或者 1 的列表，0 表示不采购，1 表示采购（约束函数 9）。$supply_d$ 为补货量列表，长度为 49，值为蔬菜单品的采购数量；$sale_d$ 为售价列表，长度为 49，值为蔬菜单品的销售单价。约束函数 1 用于满足补货量大于销售量加损耗量的要求；约束函数 2 用于满足售价大于批发价的要求；约束函数 3 用于满足最小陈列量要求；约束函数 4 用于满足单品选择数量要求；约束函数 5 用于满足补货量列表与选择列表具有相同的 0 元素值位置的要求；约束函数 6 用于满足销售价格列表与选择列表具有相同的 0 元素值位置的要求；约束函数 7 基于 4.2 所得到的蔬菜品类定价-销售量函数，通过蔬菜单品的品类和定价得到该蔬菜单品的销售量；约束函数 8 用于满足市场对各品类蔬菜商品需求，要求补货的蔬菜单品所属的品类总共有六种。求解得到商超未来 1 天的蔬菜单品补货和定价策略如下表所示：

表 4 商超未来 1 天的蔬菜单品补货和定价策略

| 单品名称 | 进货量 | 销售价格 |
| --- | --- | --- |
| 七彩椒（2） | 28.36699084 | 18.98666667 |
| 上海青 | 3.542098523 | 8 |
| 云南油麦菜（份） | 20.77066984 | 4.159060403 |
| 云南生菜（份） | 14.16476615 | 4.461504425 |
| 净藕（1） | 0 | 0 |
| 双孢菇（盒） | 0 | 0 |
| 圆茄子（2） | 28.58223041 | 6.133333333 |
| 外地茼蒿 | 0 | 0 |
| 奶白菜 | 8.539312758 | 4.792 |
| 姜蒜小米椒组合装（小份） | 20.98586099 | 4.72244898 |
| 娃娃菜 | 5.602326454 | 6.54109589 |
| 小皱皮（份） | 0 | 0 |
| 小米椒（份） | 14.09529155 | 5.769333333 |
| 小青菜（1） | 9.498576104 | 5.2 |
| 木耳菜 | 15.98918396 | 5.332075472 |

| 单品名称 | 进货量 | 销售价格 |
|---|---|---|
| 枝江青梗散花 | 0 | 0 |
| 洪湖藕带 | 0 | 0 |
| 海鲜菇（包） | 24.68027962 | 2.748387097 |
| 竹叶菜 | 12.49620554 | 3.773953488 |
| 紫茄子（2） | 0 | 0 |
| 红椒（2） | 5.627657919 | 18.89230769 |
| 红莲藕带 | 0 | 0 |
| 红薯尖 | 0 | 0 |
| 芜湖青椒（1） | 7.728847498 | 5.2 |
| 苋菜 | 0 | 0 |
| 菜心 | 0 | 0 |
| 菠菜 | 4.760075759 | 14 |
| 菠菜（份） | 4.223043276 | 5.510204082 |
| 菱角 | 29.86845451 | 14 |
| 虫草花（份） | 9.693139396 | 3.6125 |
| 螺丝椒 | 16.07547691 | 11.29133858 |
| 螺丝椒（份） | 0 | 0 |
| 蟹味菇与白玉菇双拼（盒） | 0 | 0 |
| 西兰花 | 19.3910901 | 12.4080402 |
| 西峡花菇（1） | 22.80611168 | 24 |
| 金针菇（盒） | 19.10032914 | 1.879646018 |
| 长线茄 | 0 | 0 |
| 青红杭椒组合装（份） | 29.03063805 | 5.493333333 |
| 青线椒（份） | 22.70106748 | 4.3 |
| 青茄子（1） | 4.141759794 | 6 |
| 高瓜（1） | 0 | 0 |
| 云南油麦菜 | 25.66818909 | 7.2 |
| 云南生菜 | 14.28796708 | 9.2 |
| 高瓜（2） | 0 | 0 |
| 白玉菇（袋） | 0 | 0 |
| 鲜木耳（份） | 0 | 0 |
| 木耳菜（份） | 0 | 0 |
| 紫茄子（1） | 26.75966417 | 9 |
| 野生粉藕 | 23.37169162 | 26 |

# 五、 计算结果分析及拓展

为了更好地制定蔬菜商品的补货和定价决策，我们认为商超可能还需要以下数据：

成本数据：了解每个蔬菜品类的生产和采购成本，包括生产、包装、运输和采购成本。这将有助于商超确定最佳的成本加成定价策略。

促销活动数据：收集有关促销活动的数据，包括促销周期、折扣率、促销时间段等。这有助于商超优化促销策略，并预测促销对销售的影响。

顾客反馈数据：了解顾客的反馈和购买偏好，包括商品的质量、口味、包装等方面的评价。这有助于商超改进商品选择和品质管理。

库存数据：监控库存水平，包括每个蔬菜品类的当前库存量和库存周转率。这可以帮助商超避免过多的库存或库存不足。

供应商数据：收集关于供应商的信息，包括供应商的信誉、供货稳定性和合同条款。这有助于商超选择可靠的供应商并管理供应关系。

市场趋势数据：了解蔬菜市场的趋势和变化，包括市场价格波动、新品上市、市场需求变化等。这有助于商超做出灵活的决策以适应市场变化。

综合考虑这些额外的数据，商超将能够更全面、更准确地制定补货和定价决策，以提高运营效率并满足客户需求。这些数据可用于进一步优化模型和决策，以应对多样化的市场和供应链挑战。

# 六、 模型评价

## 6.1 模型的优点

数据驱动：模型建立在大量的销售数据、商品信息和批发价格等现实数据的基础上，因此有潜力更准确地反映市场的实际需求和变化。

综合考虑：该模型考虑了多个因素，包括销售量、成本、损耗率和批发价格等，综合分析了这些因素对销售利润的影响，使决策更全面。

预测和优化：模型使用 ARIMA 模型来预测未来的批发价格，并基于非线性规划来进行优化决策，这有助于商超更好地规划补货和定价策略。

定价方法：采用了成本加成定价方法，考虑了运输损耗和品质下降对定价的影响，有助于保持合理的利润水平。

市场需求满足：在优化的过程中，模型尽量满足市场对各品类蔬菜商品的需求，这有助于提高销售额并最大化商超的收益。

## 6.2 模型的缺点

数据要求高：模型的有效性高度依赖于准确的历史销售数据、损耗率等信息。如果数据质量不高或者缺乏关键数据，模型的可行性和准确性会受到影响。

假设限制：模型中使用了一些假设，例如成本加成定价方法和 ARIMA 模型对未来价格的预测。这些假设可能不适用于所有情况，尤其是在市场波动剧烈的情况下。

计算复杂性：非线性规划模型通常需要复杂的计算和求解过程，这可能需要大量的计算资源和时间，特别是在商超的商品种类较多的情况下。

变动性：市场需求和外部因素可能会发生快速变化，例如天气、季节和竞争对手的行动等。模型可能无法及时适应这些变化。

人工干预：模型中的优化过程可能需要人工干预和调整，特别是在考虑商超的实际销售情况和需求时。这可能增加了管理的复杂性。

# 七、 总结与展望

我们在生鲜商超领域中应用了数据分析、可视化、预测和非线性规划等技术，以支持蔬菜类商品的补货和定价决策。模型的优点包括数据驱动、综合考虑多个因素、市场需求满足以及预测和优化功能。然而，模型也存在一些限制，如对高质量数据的依赖、假设的限制、计算复杂性、市场变动性和需要人工干预等。

未来可以考虑以下方向来进一步完善和发展该模型：

1. 数据质量提升：确保获得更准确和完整的历史销售数据、损耗率数据以及批发价格数据，可以通过数据清洗和采集改进来实现。

2. 模型改进：探索更复杂的模型或者机器学习方法，以更好地捕捉市场的动态变化和不确定性，特别是在需要快速决策的情况下。

3. 实时决策支持：开发实时决策支持系统，以便根据实际销售情况和市场需求进行即时调整，提高决策的灵活性。

4. 外部因素考虑：将外部因素如天气、季节和竞争对手的行动等纳入模型考虑，以更准确地预测市场变化。

5. 自动化和智能化：结合物联网（IoT）和人工智能（AI）技术，实现自动化的库存管理和定价决策，减少人工干预的需求。

6. 可扩展性：使模型具有可扩展性，可以适应不同规模和类型的商超，以满足不同的业务需求。

7. 合作与协同：考虑与供应链伙伴和批发商的合作，以优化整个供应链的效率，减少运输损耗。

总之，随着技术的不断进步和数据的积累，生鲜商超领域的决策支持模型将有更大的潜力来提高效率、减少损耗，并更好地满足市场需求。继续关注行业的发展趋势，不断改进和优化模型，将有助于实现更好的商超经营管理。

# 参考文献